%Paper: cond-mat/9505052
%From: Bertrand Berche <berche@lps.u-nancy.fr>
%Date: Fri, 12 May 95 10:26:45 WETDST

\def\d{\hbox{\rm d}}
\def\ie{\hbox{\rm i.e. }}

\hfuzz=5pt\overfullrule=0pt

\input iopppt.tex
\headline={\ifodd\pageno{\ifnum\pageno=\firstpage\titlehead
   \else\rrhead\fi}\else\lrhead\fi}

\footline={\ifnum\pageno=\firstpage
\hfil\textfonts\rm\folio\fi}
\def\titlehead{ \hfil}

\pptstyle
\jl{1}
\overfullrule=0pt

\title{Realization of supersymmetric quantum mechanics
 in inhomogeneous Ising models}[Susy
 in inhomogeneous Ising models]

\author{Bertrand Berche and Ferenc Igl\'oi\footnote\dag{Reasearch Institute for
Solid State Physics,
 P.O. Box 49,
H-1525 Budapest 114, Hungary, and Institute of Theoretical
Physics, University of Szeged, H-6720 Szeged, Hungary} }[B Berche and F
Igl\'oi]

\address{Laboratoire de Physique du Solide\footnote\ddag{Unit\'e de
Recherche Associ\'ee au CNRS No 155},  Universit\'e Henri Poincar\'e, Nancy~1,
BP 239, F--54506 Vand\oe uvre l\`es Nancy Cedex, France}

\abs
Supersymmetric quantum mechanics is well known to provide, together
with the so-called shape invariance condition, an elegant method to solve the
eigenvalue problem of some one-dimensional potentials by simple algebraic
manipulations. In the present paper, this method is used in statistical
physics.
We consider the local critical behaviour of  inhomogeneous
Ising models and  determine the complete set of anomalous dimensions from
the spectrum of the corresponding transfer matrix in the strip
geometry. For smoothly varying perturbations, the eigenvalue
problem of the transfer matrix indeed takes the form of a
Schr\"odinger equation, and the corresponding potential furthermore exhibits
the shape invariance property for some known extended defects. In
these cases, the complete spectrum is derived by the methods of
supersymmetric quantum mechanics.
\endabs

\submitted
\date

\section{Introduction}
The concept of supersymmetry has first appeared in quantum field theory and has
been later used in different areas of physics (cf random systems). The essence
of the method is more transparently seen in ordinary quantum mechanics as it is
known since the work of Witten [1]. Supersymmetric quantum mechanics [SSQM]
provides a unified framework to perform the factorization of the Schr\"odinger
equation, following the pionneering works of Dirac and
of Schr\"odinger ([2-5], for a review
see reference [6]). Furthermore, if the potential in the
Schr\"odinger
equation has the property of shape invariance [7], the eigenvalues and the
corresponding eigenvectors can be obtained by simple algebraic manipulations
and it was found that the well known
exactly solved  problems (\ie those problems
which can be rewritten as hypergeometric equations after a
suitable change of variable) exhibit the shape invariance
property..

In the present paper, we show a possible new field of application of  SSQM.  In
statistical physics, inhomogeneous systems have been extensively studied in the
past decade (for a recent review, see reference [8]). An inhomogeneity can be
caused basically in two different ways.  Geometrical effects due to the surface
shape of the system and/or modified couplings or defects may influence
 the critical behaviour. The simplest inhomogeneity is the semi-infinite
system with a free surface. The universal behaviour in a surface layer with a
width  of the order of the correlation length is described by a set of
local (surface) critical exponents which are different from the bulk ones (see
 in reference [8]). More generally, the existence of a free surface may induce
a coupling enhancement between nearest neighbour spins in a region of some
extent
close to the boundary and a local modification of the critical behaviour can
then be expected.
 One special type  of extended defect was introduced by
Hilhorst and van Leeuwen [9]. Here the couplings perpendicular to the surface
deviate from the bulk one as a power law  $A/y^\omega$, $y$ being the distance
from the free surface. It follows from a relevance-irrelevance criterion
[10,11] that this type of perturbation is marginal for the two-dimensional
Ising model at $\omega=1$.
 In this case the critical exponents are
$A$-dependent as obtained from a number of exact calculations [12-18]. The
conformal properties of such systems have been investigated using
the  plane-to-cylinder conformal mapping, under which the system is mapped
onto a
strip. Provided the perturbation profile is also properly transformed, the
gap-exponent relation [19] and the tower-like structure of the spectrum are
preserved [20-22]. This is still the case when the defect extends from a line
in
the bulk [23-27]. It was later shown, in a first order perturbation
calculation,
that the gap-exponent relation is valid for any marginal extended perturbation
[28].
On the other hand,
the geometrical shape of the free boundary may also lead to a modified critical
behaviour. These effects are relevant in the
critical behaviour at corners or parabola shaped systems (\ie such that the
boundary curve follows a parabolic law) [29-33].

In the present paper, we  consider the two-dimensional Ising model with a
marginal Hilhorst-van Leeuwen defect, as well as a related hyperbolic type
of defect in the corner geometry, and calculate the corresponding local
critical exponents.. Using conformal methods, the problems are
studied in the strip geometry. Here the spectrum of the transfer matrix  is
calculated exactly using the method of supersymmetric quantum mechanics.

The setup of the paper is the following. In section~2, we present a short
summary of SSQM
and of the concept of shape invariance of the potential partners.
 In section~3, we show that
in the cylinder geometry, when the Hamiltonian limit is considered,
   the eigenvalue equations in the
continuum limit take the form of supersymmetric Schr\"odinger
equations.
The
Hilhorst-van Leeuwen problem is considered and the complete spectrum of the
transfer matrix is calculated by the method of SSQM.  The same
calculation is performed for the hyperbolic defect in section~4.
 In
section~5,  the critical exponents are calculated and a relation between the
two problems through conformal invariance is discussed.

\section{Supersymmetric quantum mechanics}
The work of Witten [1]  has focused considerable
interest on supersymmetric quantum mechanics  (for recent
reviews see references [34,35]). Furthermore, by the concept of shape
invariance, Gendenshte\u\ii n [7] has obtained a systematic generalization to
Dirac's operator method for the $1d$ harmonic oscillator problem.

Let us consider the Hamiltonian
$$\hat {\cal H}_-=-{\d^2\over\d\zeta^2}+{\cal V}_-(\zeta
)\en$$ with a vanishing ground state energy $E^-_0$. The
ground state wave function  is then related
to the potential as ${\cal V}_-(\zeta )=\psi_0''(\zeta)/\psi_0(\zeta)$. In
terms of the superpotential
  $${\cal W}(\zeta
)=-{\d\over\d\zeta}\ln\psi_0(\zeta),\en$$
 the Hamiltonian $\hat
{\cal H}_-$ is  factorized: $$\hat {\cal
H}_-=-{\d^2\over\d\zeta^2}+\left( {\cal W}^2(\zeta)-{\cal
W}'(\zeta)\right)=\hat Q^+\hat Q^-.\en$$ Here the prime denotes
derivative with respect to $\zeta$ and the charge operators are defined
by: $$\hat Q^+=-{\d\over\d\zeta}+{\cal W}(\zeta),\qquad \hat
Q^-={\d\over\d\zeta}+{\cal W}(\zeta).\en$$

The partner Hamiltonian
$$\hat {\cal
H}_+=-{\d^2\over\d\zeta^2}+{\cal
V}_+(\zeta)=-{\d^2\over\d\zeta^2}+\left( {\cal
W}^2(\zeta)+{\cal W}'(\zeta)\right)\en$$
may then be
introduced and is also factorized, $\hat {\cal H}_+=\hat
Q^-\hat Q^+$, and there exists
a one-to-one correspondence between the spectrum of the two partner
Hamiltonians as: $E^-_{n+1}=E^+_n$. If the ground state wave functions of $\hat
{\cal H}_\pm$, which are given by equation~(2) as $$\psi_0^\pm (\zeta
)=\exp\left[ \pm\int {\cal W}(\zeta)\d \zeta\right]\en$$
are non normalizable, the ground state energies of
both $\hat {\cal H}_-$ and $\hat {\cal H}_+$ are non-zero
and $E^-_n=E^+_n$. In this  case supersymmetry is broken.

In the following we consider unbroken supersymmetry, \ie $E_0^-=0$ and the
potential partners which satisfy the shape invariance property as:
 $${\cal
V}_+(\zeta,a_0)={\cal V}_-(\zeta,a_1)+R(a_1).\en$$
Here $a_0$ is a parameter
of the Hamiltonian, $a_1$ is some function of $a_0$, and $R(a_1)$ is a function
which does not involve the variable $\zeta$. It is then easy to show
that the  spectrum of $\hat {\cal H}_-$ and $\hat {\cal
H}_+$ are simply shifted by the amount of $R(a_1)$ and
then, by iterating   the shape invariance relation, one builds a hierarchy
of Hamiltonians whose spectra are related as mentioned above. Finally, one
finds  the
 eigenvalues of $\hat {\cal H}_-$ as:
$$E^-_n(a_0)=\sum_{k=1}^nR(a_k).\en$$
The corresponding wave functions are  obtained by applying the charge
operators on the ground state wave function: $$\psi_n(\zeta ,a_0)\sim \hat
Q^+(a_0)\hat Q^+(a_1)\dots\hat Q^+(a_{n-1})\psi_0(\zeta,a_n).\en$$

The shape invariant potentials can be found in the literature [36-41].
The
factorization technique was in fact originally introduced in the context of
ordinary differential equations by Darboux [42-44], and the application of the
so-called commutation formula to the Schr\"odinger equation can already be
found
in [45].

\section{Hilhorst-van Leeuwen model}
Consider a semi-infinite two dimensional Ising model with inhomogeneous nearest
neighbour couplings
$$K(\rho ,\theta )=K(\infty )-g{\cal Z}(\rho ,\theta )\en$$
where $K(\infty )$ is the bulk critical value. The scale covariance requirement
for the inhomogeneity leads  to a
power-law behaviour for the radial part of the shape function
[28]:
$${\cal Z}(\rho ,\theta )=f (\theta )/\rho^\omega
,\en$$
and the perturbation amplitude $g$, then, scales
under renormalization as  $g'=b^{y_t-\omega}g$
where $y_t$ is the bulk thermal exponent. Here, we use the method of conformal
invariance. The deviation from the bulk
coupling in the original system $t(z)=K(\rho,\theta)-K(\infty)$ transforms,
under the conformal mapping
$w=w(z)=u+iv$,
according to $t(w)=\mid w'(z)\mid^{-y_t}t(z)$ [20]. With the usual
plane-to-cylinder logarithmic conformal mapping $w(z)={L\over\pi}\ln
z$, the semi-infinite system is mapped onto an infinitely long strip of width
$L$ with free boundary conditions, and the inhomogeneity (10) becomes
$$K(u,v)=K'(\infty )-g\left( {\pi\over L}\right)^\omega\exp\left[ {\pi
u\over L}(\omega-y_t )\right] f\left({\pi v\over L}\right) ,\en$$
where $K'(\infty)$ is the critical coupling in the modified geometry.
If we furthermore assume a marginal inhomogeneity,  \ie such that
the perturbation amplitude remains unchanged under a rescaling, one has
$\omega=y_t$ and it yields a perturbation which is
independent of the $u-$direction along the strip:
$$K(v)=K'(\infty )-g{\pi\over L} f\left( {\pi v\over
L}\right) .\en$$

The prototype of smoothly inhomogeneous systems has been introduced
by
Hilhorst and van Leeuwen [9]. Here, as an illustration, we recover the
results previously obtained
by Burkhardt and
Igl\'oi [20] by more complicated methods.
Consider a two-dimensional semi-infinite Ising model on a  square
lattice. The couplings $K_1$ parallel to the surface are
constant, while the nearest neighbour couplings $K_2(y)$
perpendicular to the surface assume a power law deviation from
their bulk critical value (figure~(1a)):
$$K_2(y)=K_2(\infty )-{g\over y},\en$$
where $y$ measures the distance from the free surface. This
corresponds to a marginal shape
function ${\cal Z}(\rho,\theta)=(\rho\sin\theta)^{-1}$. This model has  been
extensively studied in the two-dimensional classical
version [9-16] as well as in its quantum counterpart [17,18,20-23]
(for a review see reference [8]).
 Following Burkhardt and Igl\'oi [20], we
transform the inhomogeneity by the logarithmic conformal mapping and the
inhomogeneity transforms into a sinusoidal form on the strip:
$$K_2(v)=K'_2(\infty)-{\pi\over L}{g\over\sin\left( {\pi v\over
L}\right)}\qquad
0<v<L.\en$$ The transfer matrix along the strip $\hat {\cal
T}=e^{-\tau\hat{\cal
H}}$, where $\tau$ is the lattice spacing, leads, in the extreme anisotropic
limit [46-48], to a one-dimensional quantum chain defined by the Hamiltonian:
$$\hat{\cal H}=-{1\over 2}\sum_{\ell=1}^L\sigma_z (\ell) - {1\over 2}\sum_{\ell
=1}^{L-1}\lambda (\ell)\sigma_x(\ell)\sigma_x(\ell +1),\en$$ with varying
couplings $$\lambda(\ell)=1-{\pi\over L}{\alpha\over\sin\left({\pi\ell\over
L}\right)}.\en$$ Here, the $\sigma$'s are the Pauli matrices. The Hamiltonian
$\hat{\cal H}$ can be diagonalized by standard methods [49,50], transforming in
terms of fermion creation ($\eta^+_k$) and annihilation ($\eta_k$) operators as
$$\hat{\cal H}=\sum_k\Lambda_k\left(\eta_k^+\eta_k-{1\over 2}\right) .\en$$
Here, the fermionic modes with the lowest energies, which are $\Or(L^{-1})$,
are
obtained in the continuum approximation from a pair of Schr\"odinger equations
involving the inhomogeneity
  function $\chi (\zeta )=\alpha /\sin\zeta$ where $\zeta=\pi\ell /L$.
The first one in terms of $\psi_k$ reads as
 $$-{\d ^2\psi_k\over\d
\zeta^2}+\left(\chi^2(\zeta )-\chi '(\zeta
)\right)\psi_k(\zeta )=\left( {\Lambda_k
L\over\pi}\right)^2\psi_k(\zeta ),\quad
0\leq\zeta\leq\pi
\enpt$$
with the boundary conditions
$$\left.\psi_k(\zeta)\right|_{\zeta =0} =0,\quad
\left.{\psi_k'(\zeta
)\over\psi_k(\zeta)}\right|_{\zeta=\pi}=-\chi (\pi
).
\endpt$$
Similarly for the function $\phi_k$:
 $$-{\d ^2\phi_k\over\d
\zeta^2}+\left(\chi^2(\zeta )+\chi '(\zeta
)\right)\phi_k(\zeta )=\left( {\Lambda_k
L\over\pi}\right)^2\phi_k(\zeta ),\quad
0\leq\zeta\leq\pi ,
\enpt$$
$$
\left.{\phi_k'(\zeta
)\over\phi_k(\zeta)}\right|_{\zeta=0}=+\chi (0),\quad
\left.\phi_k(\zeta)\right|_{\zeta =\pi} =0 .
\endpt$$
In these expressions, $\psi(\zeta)$ and $\phi(\zeta)$ are the
continuum limit approximations of
 the eigenvectors entering the discrete eigenvalue equations that one
obtains when
diagonalizing the Hamiltonian (16) (see reference [49]).

The similarity between these equations and the  Schr\"odinger
equations  encountered in supersymmetric quantum
mechanics has been already mentioned by Choi [51], but here we show
how the concept of
shape invariance may  be used  to determine the excitation
spectrum.

First, we note that all the eigenvalues of equations~(19a) and (20a)
are the same,
including the smallest one $(\Lambda_0 L/\pi
)^2$, thus, in the language of SSQM, supersymmetry is broken
[52,53]. This statement is in agreement with Witten's argument [1],
according to which
unbroken supersymmetry requires a superpotential with one node (or an odd
number of nodes). This is obviously not the case for $\chi (\zeta)$ (see
figure~2), which is symmetrical to $\pi /2$, since the inhomogeneity
in the semi-infinite plane is translationally invariant along the
surface. We have then to face the problem of finding a superpotential ${\cal
W}(\zeta) $ in order to restore supersymmetry. This superpotential must be
related to the inhomogeneity function $\chi (\zeta )$ by a Riccati equation,
\ie such that ${\cal W}^2-{\cal W}'$ and $\chi^2-\chi '$ are identical up
to a constant, the constant being of essential importance because its existence
ensures that supersymmetry will be restored.

The boundary conditions in equations~(19b) and (20b) generally pose
the same requirement as in SSQM, \ie the wavefunction must
vanish at both ends of the interval since the potential term
diverges there. However,
if $\psi'$ diverges faster than $\psi$ when
$\zeta\rightarrow\pi$,
then the solution of equation~(19a) is non normalizable. This type of
solution, which describes a localized mode, is associated to the
appearence of spontaneous surface order in the system and
corresponds to a vanishing excitation $\Lambda_0=0$. For the
Hilhorst-van Leeuwen inhomogeneity, such a solution is given by
$$\psi_{\hbox{\sevenrm loc.}}(\zeta)\sim\exp\left( -\int\chi (\zeta
)\d\zeta\right)=\tan^{-\alpha}\left({\zeta\over
2}\right),\en$$
which is indeed non  normalizable for $\alpha<-{1\over 2}$. The
lowest excitation energy in this region is then $\Lambda_0=0$. We
shall return later to determine the higher lying levels of the
spectrum in this case.

In the following, we deal with the region $\alpha\geq-{1\over 2}$,
where the method of SSQM works without limitations and  shape invariance
is a worthwhile concept to deduce the eigenvalue spectrum.
First, we should  find
a convenient superpotential
which solves the Riccati equation.
This is done with the mapping introduced by Dutt et al [54]. The inhomogeneity
function $\chi(\zeta)$ is a special case of the Scarf superpotential
$S(\zeta)=\alpha_1/\sin\zeta-\alpha_2\cot\zeta$, which leads to  the
Eckart potential by $S^2(\zeta)-S'(\zeta)$ [55]. With the choice
$\alpha_1=-1/2$
and $\alpha_2=\alpha+1/2$, this defines a new superpotential
$${\cal
W}_>(\zeta)=-{1\over 2\sin\zeta}-\left(\alpha +{1\over 2}\right)\cot\zeta.
\en$$
 It
is also easy to see that the superpotential ${\cal }W_>(\zeta)$
presents one node (figure~2), thus Witten's requirement on unbroken
supersymmetry is satisfied and supersymmetry
is now unbroken in the range $\alpha\geq-{1\over 2}$.
This choice  leads to
  the trigonometric Eckart potential
 for ${\cal V}_-(\zeta )$:
 $${\cal V}_-(\zeta)={\alpha^2+\alpha\cos\zeta\over\sin^2\zeta}
-\left(\alpha +{1\over 2}\right)^2,\en$$
and the  ground state excitation
$\Lambda_0$ can thus be identified as:
$$\left({\Lambda_0L\over\pi}\right)^2=\left(\alpha +{1\over
2}\right)^2.\en$$
 Now equation~(19a) can be written as:
$$\fl -{\d ^2\psi_k\over\d \zeta^2}+\left({\cal W}_>^2(\zeta
)-{\cal W}_>'(\zeta )\right)\psi_k(\zeta )=\left[\left(
{\Lambda_k L\over\pi}\right)^2-
\left(
{\Lambda_0 L\over\pi}\right)^2\right]
\psi_k(\zeta ).\en$$
and the  ground state
wave function, obtained through equation (6) is given by:
$$\psi_0^>(\zeta)\sim\exp\left( -\int{\cal W}_>(\zeta )\d\zeta\right)
\sim{1\over\sqrt\pi}\sin^{\alpha +1}\zeta \ (1+\cos\zeta )^{-1/2}.\en$$
The solution (26), which is  indeed normalizable for $\alpha\geq -1/2$,
  continuously
evolves towards the localized mode (21) when
$\alpha\rightarrow\alpha_c=-{1\over 2}$ from above.
The higher lying levels of the Schr\"odinger equation, which are
given as
$$E_k^-=\left[\left(
{\Lambda_k L\over\pi}\right)^2- \left( {\Lambda_0
L\over\pi}\right)^2\right],\en$$
 are  obtained from the shape invariance
property of the partner potentials: $${\cal V}_+(\zeta, a_0)={\cal
V}_-(\zeta, a_1)+\left( a_1+{1\over 2}\right)^2-\left(
a_0+{1\over 2}\right)^2\en$$
where $a_0=\alpha$, $a_1=a_0+1$.
Then, according to equations (7) and (8), the energies of the
single fermion excitations follow  from the remainder function
$R(a_1)=(a_1+1/2)^2-(a_0+1/2)^2$:
$$\Lambda_k={\pi\over L}\left(\alpha +k+{1\over
2}\right),\qquad
k=0,1,2\dots ,\qquad\alpha\geq -{1\over 2}.\en$$

In the  regime of surface order, $\alpha <-1/2$,
the eigenfunctions of the excited states of equation~(19a) are
normalizable,  the previous method thus
applies. Now,  the superpotential is  given by
$${\cal W}_<(\zeta)={1\over 2\sin\zeta}+\left(\alpha -{1\over
2}\right)\cot\zeta \en$$
and the energy of the first non-vanishing excitation is  identified
as:
  $$\left({\Lambda_1L\over\pi}\right)^2=\left({1\over
2}-\alpha \right)^2.\en$$
The higher lying excitations can be
similarly obtained from the shape invariance property, so that the
energies of the fermion modes are now given  as:
$$\fl\Lambda_0=0,\qquad\Lambda_{k}={\pi\over
L}\left(k-\alpha-{1\over 2} \right),\qquad
k=1,2,3\dots ,\qquad\alpha\leq -{1\over 2}.\en$$

The potential ${\cal V}_-(\zeta)$ and the corresponding eigenenergies $E_k^-$
are
shown in figure~3 in the ordered phase ($\alpha<\alpha_c$) and in the
non-ordered phase ($\alpha>\alpha_c$). Figure~4 shows the two first
eigenfunctions in the two regimes.

\section{Hyperbolic defect}
The inhomogeneity in the Hilhorst-van Leeuwen model, studied before,
can be considered as a result of elastic deformations on the free
surface of the system. If the system has now the shape of a corner
with  a right angle, then uniform elastic deformations would result
in a defect of hyperbolic form. The local couplings are constant along the
hyperbolas
$f(x,y)={1\over xy}=const$,
whereas the couplings pointing perpendicular to the $f(x,y)$ lines
are assumed to vary as (figure~(1b)):
$$K_\perp(x,y)=K_\perp (\infty )-g{\rho\over xy}.\en$$
Thus, near to the surface but far from the corner, the
inhomogeneity has the same shape as in the Hilhorst-van Leeuwen
model in equation~(10).
The shape
function corresponding to this defect is:
$${\cal Z}(\rho,\theta )={1\over\rho\cos\theta\sin
\theta}.\en$$
Once again, the inhomogeneity is mapped onto a
strip geometry, now the appropriate conformal transformation is the the
 ${2L\over\pi}\ln z$ logarithmic mapping. In the strip geometry,
the inhomogeneity is again  a sinusoidal form, the couplings in the
Hamiltonian operator (16) vary as:
$$\lambda (\ell )=1-{\pi\over2 L}{\alpha\over
\displaystyle\cos\left({\pi\ell\over2 L}\right)
\sin\left({\pi\ell\over 2L}\right)}=1-{\pi\over
L}{\alpha\over\displaystyle\sin\left( {\pi\ell\over L}\right)}\en$$
from which we
deduce the inhomogeneity function in the continuum limit:
$$\chi(\zeta )={\alpha \over\cos\zeta\sin \zeta}={2\alpha\over\sin
2\zeta},\en$$
where $\zeta$ is defined in the range $0\leq
\zeta\leq\pi/2$. This inhomogeneity function leads to the P\"oschl-Teller
potential, but here, from the previous section we can  immediately get the
energies of the single particle excitations:
$$\Lambda_k={\pi\over 2L} (2\alpha +2k+1),\qquad
k=0,1,2\dots ,\qquad\alpha\geq -{1\over 2},\enpt$$
$$\Lambda_0=0,\quad\Lambda_k={\pi\over 2L} (2k-1-2\alpha),\quad
k=1,2,3\dots ,\quad\alpha\leq -{1\over 2}.\endpt$$

\section{Local critical properties}
Conformal invariance makes it possible to transform critical systems from one
restricted geometry into another, and deduce the local critical exponents in
the
former geometry from the energy gaps in the transformed one.

In a semi-infinite system, like the Hilhorst-van
Leeuwen model, the algebraic decay of the correlation function at the critical
point between one point close to the surface ($z_1\sim 1$) and another point
far
in the bulk ($z_2\sim z$) is asymptotically given as $\mid
z\mid^{-(x+x_1^\mu)}$, where $x_1^\mu$ is the surface anomalous dimension of
the operator $\mu$, while $x$ is the corresponding exponent for the homogeneous
bulk. Under the logarithmic conformal mapping $w(z)={L\over\pi}\ln z=u+iv$, the
correlation function transforms according to the usual position-dependent law
involving only the bulk scaling dimensions:
$$\langle\mu(w_1)\mu(w_2)\rangle=\mid w'(z_1)\mid^{-x}\mid w'(z_2)\mid^{-x}
\langle\mu(z_1)\mu(z_2)\rangle\en$$ and the correlations in the cylinder
geometry exhibit an exponential decay along the strip which defines
the correlation length $\xi$ on the strip. In the extreme anisotropic limit,
$1/\xi$ is given by the energy gap [47], so that  the surface anomalous
dimensions are contained in the spectrum of the Hamiltonian operator in
equation~(16): $$x_1^\mu={L\over\pi}(E_\mu-E_0)\en$$
as $L\rightarrow\infty$. Then, using the diagonal form of $\hat{\cal H}$ in
equation~(18), the surface critical exponents of the Hilhorst-van Leeuwen
model can be obtained as combinations of the $\Lambda_k$ fermion energies. For
example the critical exponents of the surface magnetization and surface energy
correlations are given by:
$$x_1^m=\alpha +{1\over 2},\quad x_1^e=2\alpha+2,\qquad\alpha\geq
-{1\over 2},\enpt$$
$$x_1^m=0,\quad x_1^e={1\over 2}-\alpha,\qquad\alpha\leq
-{1\over 2},\endpt$$
in agreement with [20]. Here, $x_1^m=0$ is due to surface ordering.

For the hyperbolic defect, one defines the corner exponents, denoted $x_c^\mu$,
and associated to the algebraic decay of correlations in the corner geometry.
In the strip geometry, the $x_c^\mu$'s are again proportional to the
corresponding gaps of the Hamiltonian operator such as
$$x_c^\mu={L\over\Theta}(E_\mu-E_0).\en$$
Then, the corner exponents for the magnetization and the energy for the
hyperbolic defect with a right angle $\Theta=\pi/2$ are given as:
$$x_c^m=2\alpha +1,\quad x_c^e=4\alpha+4,\qquad\alpha\geq
-{1\over 2},\enpt$$
$$x_c^m=0,\quad x_c^e=1-2\alpha,\qquad\alpha\leq
-{1\over 2}.\endpt$$

Comparing these results to those of the Hilhorst-van Leeuwen model, one can
notice that the corner exponents are in each case the double of the
corresponding surface ones.
The same relation is known between exponents at a free surface and those of a
corner of a right angle without the presence of an inhomogeneity, which is,
according to Cardy [33], a consequence of conformal invariance. The Schwarz
mapping
$\tilde z=z^{\Theta /\pi}$ with $\Theta=\pi/2$ connects the two geometries and
leads to the above relation between the local exponents. It is not difficult to
see that the same Schwarz mapping transforms the Hilhorst-van Leeuwen
inhomogeneity and the hyperbolic defect into each other, and thus gives the
explanation for the observed relation between the corresponding local scaling
dimensions. This last result can be used in the opposite direction, then the
close relation between the spectrum of the Eckart and that of the
P\"oschl-Teller
potentials can be attributed to conformal symmetry.

Finally we note that the relation between SSQM and inhomogeneous Ising models
cannot be exploited further. Inspecting the table of shape invariant
superpotentials [36-41], no further one is known at present which could serve
as a basis for a new physically relevant inhomogeneity with an exact solution
on the two-dimensional Ising model.

\ack We thank L. Turban for valuable discussions. This work has been supported
by the CNRS and the Hungarian Academy of Sciences through an exchange program.
The work of F.I. has been supported by the Hungarian National Research Fund
under Grant No OTKA T012830.

\references
\numrefjl{[1]}{Witten E 1981}{Nucl. Phys. B}{185}{513}
\numrefbk{[2]}{Dirac P A M 1958}{The principles of
quantum mechanics}{(Oxford: Oxford University Press)}
\numrefjl{[3]}{Schr\"odinger E 1940}{Proc. R. Irish
Acad.}{A46}{9}
\numrefjl{[4]}{\dash 1940}{Proc. R. Irish
Acad.}{A46}{183}
\numrefjl{[5]}{\dash  1941}{Proc. R. Irish
Acad.}{A47}{53}
\numrefjl{[6]}{Infeld L and Hull T E 1951}{Rev. Mod. Phys.}{23}{21}
\numrefjl{[7]}{Gendenshte\u\ii n L \'E 1983}{JETP Lett.}{38}{356}
\numrefjl{[8]}{Igl\'oi F, Peschel I and Turban L 1993}{Adv.
Phys.}{42}{683}
\numrefjl{[9]}{Hilhorst H J and van Leeuwen J M J 1981}{Phys.
Rev. Lett.}{47}{1188}
\numrefjl{[10]}{Burkhardt T W 1982}{Phys. Rev. B}{25}{7048}
\numrefjl{[11]}{Cordery R 1982}{\PR B}{48}{215}
\numrefjl{[12]}{Burkhardt T W and Guim I 1982}{\JPA}{15}{L305}
\numrefjl{[13]}{Bl\"ote H W J and Hilhorst H J 1983}{Phys. Rev.
Lett.}{51}{20}
\numrefjl{[14]}{Burkhardt T W and Guim I 1984}{Phys. Rev.
B}{29}{508}
\numrefjl{[15]}{Burkhardt T W, Guim I, Hilhorst H J and van
Leeuwen J M J 1984}{Phys. Rev. B}{30}{1486}
\numrefjl{[16]}{Bl\"ote H W J and Hilhorst H J
1985}{\JPA}{18}{3039}
\numrefjl{[17]}{Peschel I 1984}{Phys. Rev. B}{30}{6783}
\numrefjl{[18]}{Kaiser C and Peschel I 1989}{J. Stat.
Phys.}{54}{567}
\numrefjl{[19]}{Cardy J L 1984}{\JPA}{17}{L385}
\numrefjl{[20]}{Burkhardt T W and Igl\'oi F 1990}{\JPA}{23}{L633}
\numrefjl{[21]}{Igl\'oi F 1990}{Phys. Rev. Lett.}{64}{3035}
\numrefjl{[22]}{Berche B and Turban L 1990}{\JPA}{23}{3029}
\numrefjl{[23]} {Igl\'oi F, Berche B and Turban L 1990}{Phys. Rev.
Lett.}{65}{1773}
\numrefjl{[24]}{Bariev R Z 1988}{Zh. Eksp. Teor. Fiz.}{94}{374 (1988 Sov.
Phys.-JETP {\bf 67} 2170)}
\numrefjl{[25]}{\dash 1989}{\JPA}{22}{L397}
\numrefjl{[26]}{Bariev R Z and Malov O A  1989}{Phys. Lett.}{136A}{291}
\numrefjl{[27]}{Bariev R Z and Ilaldinov I Z 1989}{\JPA}{22}{L879}
\numrefjl{[28]}{Turban L and Berche B 1993}{\JPA}{26}{3131}
\numrefjl{[29]}{Peschel I, Turban L and Igl\'oi F 1991}{\JPA}{24}{L1229}
\numrefjl{[30]}{Cardy J L 1983}{\JPA}{16}{3617}
\numrefjl{[31]}{Barber M N, Peschel I and Pearce  P A 1984}{J. Stat.
Phys.}{37}{497}
\numrefjl{[32]}{Davies B and Peschel I 1991}{\JPA}{24}{1293}
\numrefjl{[33]}{Cardy J L 1984}{Nucl. Phys. B}{240}{[FS12] 514}
\numrefjl{[34]}{Gendenshte\u\ii n L \'E and Krive I V 1985}{Sov. Phys.
Usp.}{28}{645}
\numrefbk{[35]}{Cooper F, Khare A and Sukhatme U P
1994}{preprint}{LA-UR-94-569}
\numrefjl{[36]}{Dabrowska J W, Khare A and Sukhatme U P
1988}{\JPA}{21}{L195}
\numrefjl{[37]}{Cooper F, Ginocchio J N and Khare A
1987}{Phys. Rev. D}{36}{2458}
\numrefjl{[38]}{Dutt R, Khare A and Sukhatme U P 1988}{Am. J.
Phys.}{56}{163}
\numrefjl{[39]}{L\'evai G 1989}{\JPA}{22}{689}
\numrefjl{[40]}{\dash 1992}{\JPA}{25}{L521}
\numrefjl{[41]}{Chuan C X 1991}{\JPA}{24}{L1165}
\numrefjl{[42]}{Darboux G 1882}{C. R. Acad. Sci. (Paris)}{94}{1456}
\numrefjl{[43]}{Andrianov A A, Borisov N V and Ioffe M V 1985}
{Theor. Math. Phys. (USSR)}{61}{1078}
\numrefjl{[44]}{Reach M 1988}{Commun. Math. Phys.}{119}{385}
\numrefjl{[45]}{Deift P A 1978}{Duke Math. J.}{45}{267}
\numrefjl{[46]}{Fradkin E and Susskind L 1978}{\PR D}{17}{2637}
\numrefjl{[47]}{Kogut J B 1979}{Rev. Mod. Phys.}{51}{659}
\numrefbk{[48]}{Christe P and Henkel M 1993}{Introduction to Conformal
Invariance and Its Applications to Critical
Phenomena}{(Berlin, Heidelberg: Springer-Verlag)}
\numrefjl{[49]}{Lieb E H, Schultz T D and Mattis D C 1961}{Ann.
Phys. (N. Y.)}{16}{406}
\numrefjl{[50]}{Pfeuty P 1970}{Ann. Phys. (N. Y.)}{57}{79}
\numrefjl{[51]}{Choi J Y 1993}{\JPA}{26}{L331}
\numrefjl{[52]}{Sukumar C V 1985}{\JPA}{18}{2917}
\numrefjl{[53]}{Lancaster D 1984}{\NC}{79}{28}
\numrefjl{[54]}{Dutt R, Gangopadhyaya A, Khare A, Pagnamenta A and
Sukhatme U 1993}{Phys. Lett. A}{174}{363}
\numrefjl{[55]}{Eckart C 1930}{\PR}{35}{1303}

\Figures

\figure{Enhancement of local couplings near an extended defect,  a) at a free
surface (Hilhorst-van Leeuwen inhomogeneity), b) at a corner
(hyperbolic defect).}
\figure{Inhomogeneity function $\chi(\zeta)$
(\dashed) and superpotential ${\cal W}(\zeta)$  (\full) for the Hilhorst-van
Leeuwen model for  $\alpha=2$ (left) and $\alpha=-4$ (right).}
\figure{${\cal V}_-(\zeta)$ potential (\full) and allowed
eigenenergy levels $E_k^-$ (\dashed) for
$\alpha=2$ (left) and $\alpha=-4$ (right)}
\figure{
Ground state  and first excited wave functions for the Hilhorst-van Leeuwen
model for
$\alpha=2$ (left) and $\alpha=-4$ (right).}

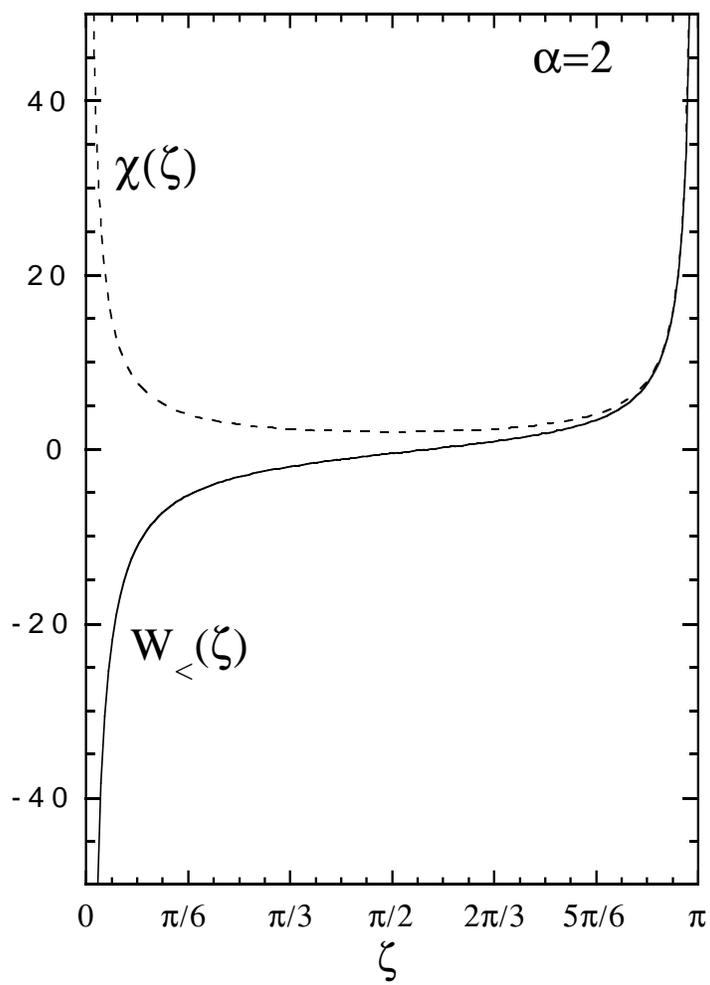
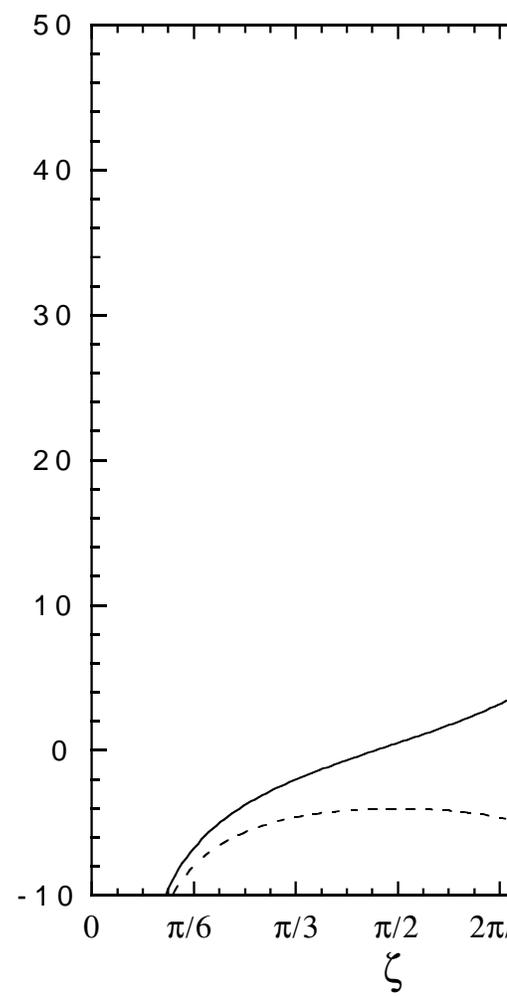
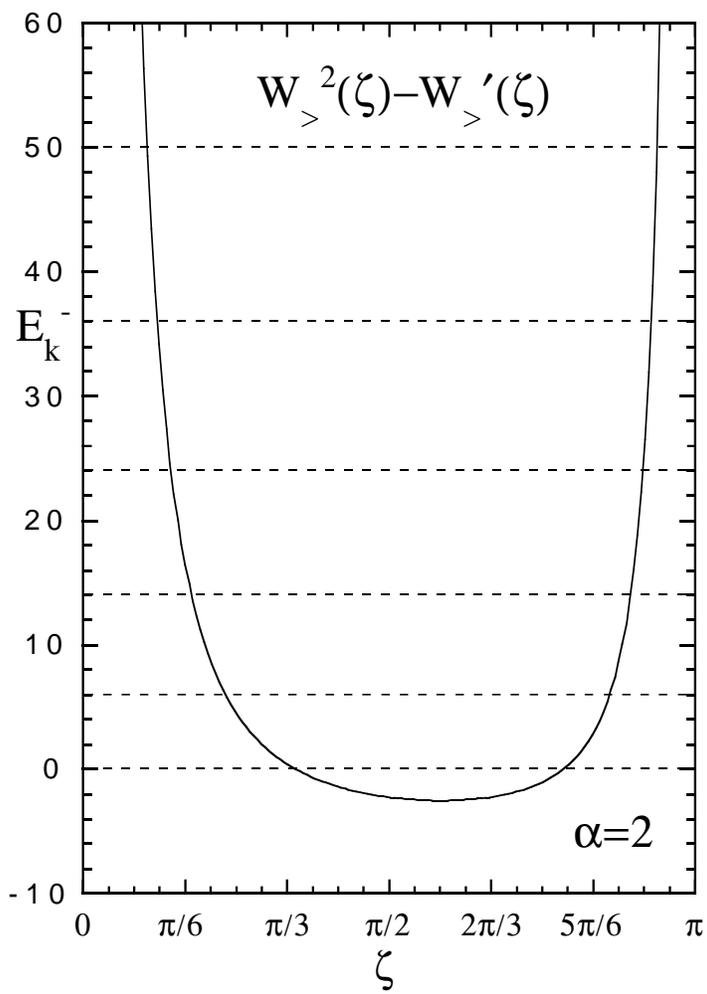
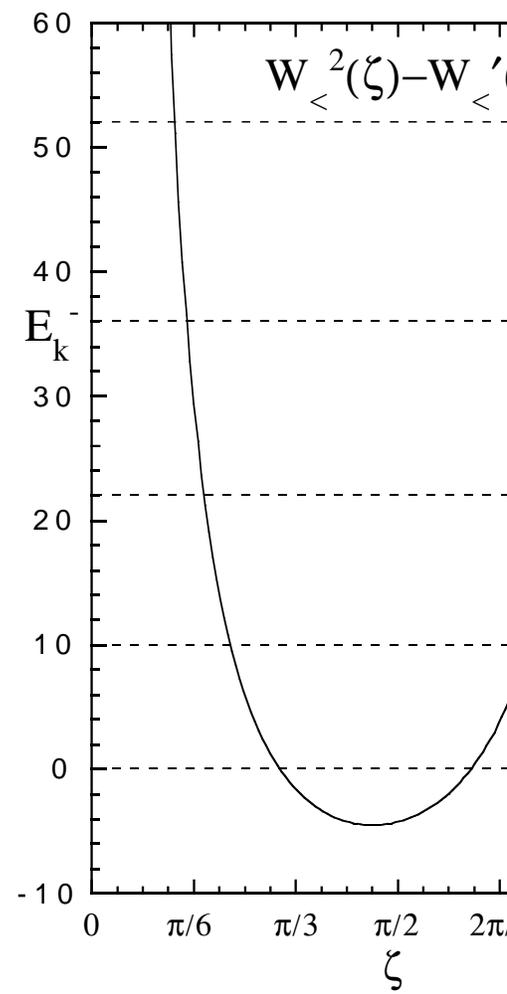
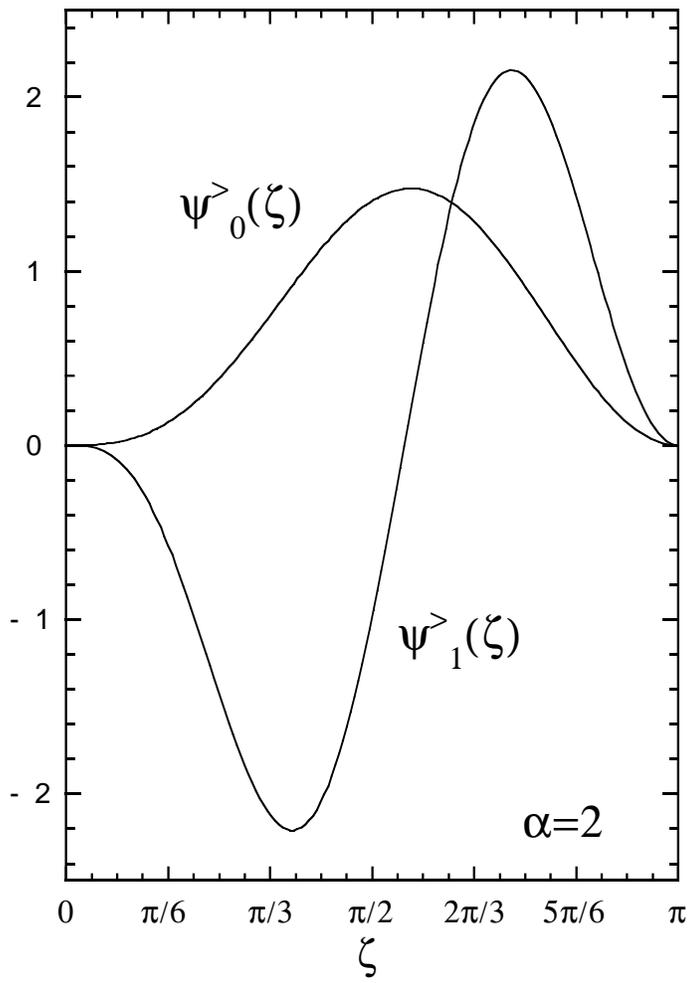
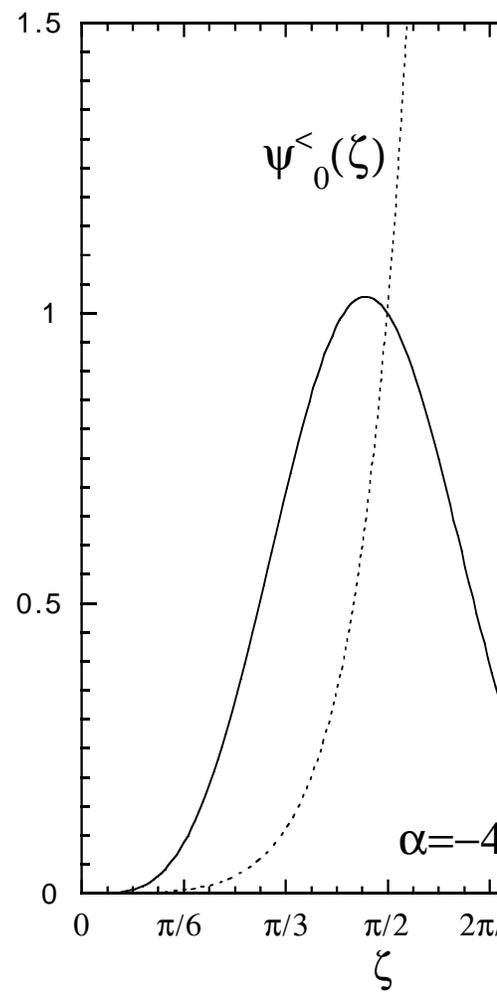
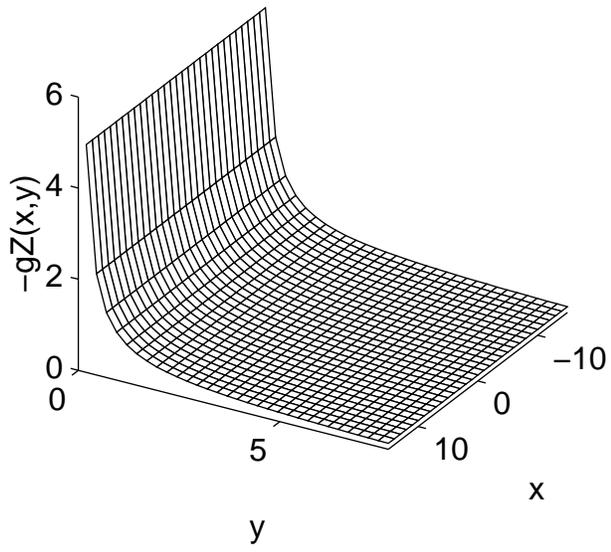
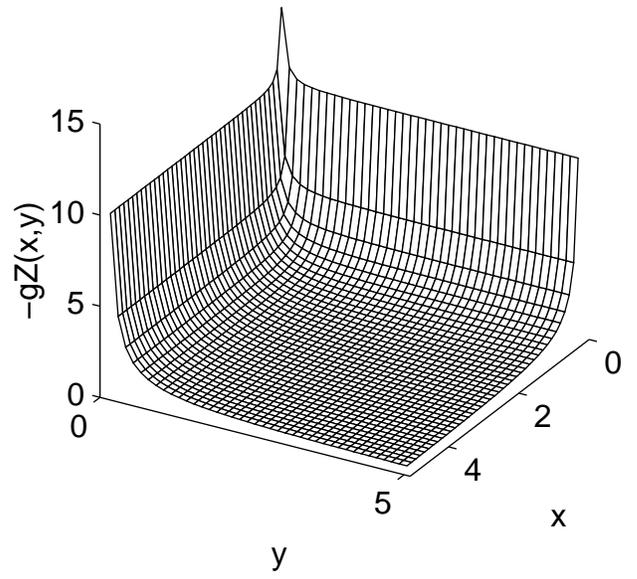
\bye